# INTEGRATIVE WIRELESS DEVICE FOR REMOTE CONTINUOUS BLOOD BIOMARKER MONITORING

**Dinh-Tuan Phan\*, Kerwin Kwek Zeming, Sophie Wan Mei Lian, Lin Jin, Ngoc-Duy Dinh, and Chia-Hung Chen\***

[1]Department of Biomedical Engineering, National University of Singapore, SINGAPORE
[2]Department of Biomedical Engineering, City University of Hong Kong, HONG KONG

**ABSTRACT**

To perform precision medicine in real-time at home, a device capable of long-distance continuously monitoring target biomolecules in unprocessed blood under dynamic situations is essential. In this study, an integrative buffer-free wireless device is developed to measure drug concentrations in patient's blood in real time for remote clinical healthcare. To demonstrate its capability, the drug molecules (i.e., small-molecule drug doxorubicin, DOX) are continuously measured in the unprocessed whole blood of live animals (e.g., rats). The dynamic changes of drug concentrations with sub-minute temporal resolution are recorded for an extended period of time (~8 hours). As an advance in remote diagnosis, this device would benefit the public by enabling long-distance precision medicine to prevent pandemics in advance.

**KEYWORDS:** Wireless healthcare, Real time bioassay, Buffer-free device, Whole bloods tests

**INTRODUCTION**

The remote continuous monitoring of health conditions could revolutionize healthcare through the ability to administer the right drug at the right time[1]. With the capability of precision medicine at home, the decentralization of the medical system would be approached to avoid unnecessary infections. However, because of the complication in sample extractions, washing, and detections, only limited physiological indexes (e.g., heartbeat, blood pressure) could be measured by using current wireless biomedical devices. Indeed, although there are a range of important biomarkers in bloods[2], the challenge to remotely continuously monitor blood samples at home remains. In this study, we develop an all in one integrated wireless device to extract and measure the biomarker concentrations (drugs) in blood without any pre-processing steps (e.g., blood dilution) for continuous measurement. With the capability of continuous monitoring of blood samples, an e-aptamer receptor is used to record the electrochemical signals of binding biomarkers in real time by a mobile device to monitor the dynamic changes of drug concentrations with the sub-minute temporal resolution for an extended period of time (i.e., up to several hours).

**EXPERIMENTAL**

To fabricate a wireless device for blood tests, in the first step, the plasma is extracted from whole blood by using L-shape Deterministic Lateral Displacement (DLD)[3] microarray and microfluidic reservoirs. After that, the drug molecules in plasma are measured via the aptamer sensors to show the wireless electrical signals, recorded by a mobile phone. After assaying, the ion concentration polarization (ICP)[4-7] is triggered via an electrical nanofluidic component to wash out the bound aptamers without uploading additional clean buffers for automatic signal regeneration for the next time point measurement. The target drug molecules are measured via aptamers to show the electrical signals. The electrical signals are then transferred to a wireless component for drug concentration monitoring by using a hand-held wireless potentiostat coupled with a mobile phone. After each measurement, the drug molecules can be automatically removed from aptamers by triggering the ICP phenomenon for signal regeneration. Thus, the aptamers can be ready for the subsequent measurements.

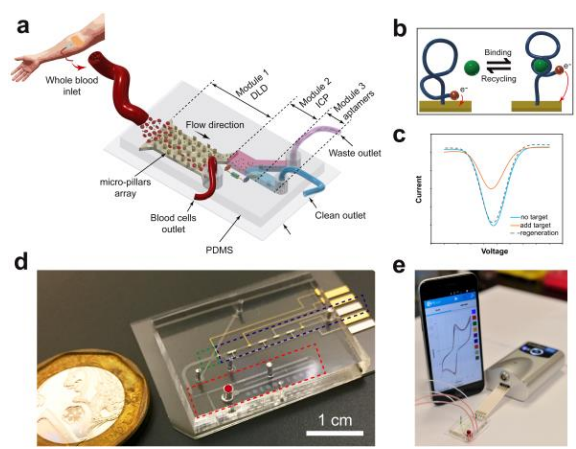

*Figure 1: Integrative wireless device working principle.*

**RESULTS AND DISCUSSION**

An integrated wireless device is fabricated for whole blood real-time biomarker (drug) monitoring. The device is constructed by four parts: 1. plasma extraction component (Deterministic Lateral Displacement, DLD microarray), 2. microfluidic reservoir, 3. wireless aptamer sensors, and 4. ICP nanofluidic component. Two-stage DLD device is designed to completely remove the RBCs and ~ 90% platelets in whole blood. At first, a large amount of RBCs is removed from unprocessed whole blood by cell sedimentation method with two reservoirs located at the inlet. The first reservoir is placed on the upper PDMS layer for sample loading. The second microfluidic reservoir is for cell sedimentation. After that, the plasma is further refined by two L-shape DLD pillars arrays fabricated with a cut-off size Dc = 2.4 µm. With two L-shaped DLD arrays connected in serial, the red blood cells can be removed completely without uploading buffers for dilution. After removing red blood cells, ~ 90 % of platelets, and random particles, the refined plasma flows to the aptamer sensing part to quantify the drug concentrations via aptamer binding. After the measurement, the ICP is triggered by applying an electrical field to convert the plasma to be clean solution for washing. After ~3-5 minutes washing, the aptamers change back to their initial (unbound) states for signal regeneration. The device with a standard electrical connection is connected to a commercial wireless potentiostat for measuring the electrical signals for remote monitoring. To perform the animal experiments, 6 male Sprague Dawley rats are procured from INVIVOS. The rats are placed on a heating pad and anesthetized during the drug injections and collections. A negative pressure was created in the syringe, tail vein located, and the catheter was inserted. 50 µL of blood is withdrawn and collected in a tube with 30 µL of heparin. The signals are then recorded by using the device to monitor the drug concentrations *in vivo*.

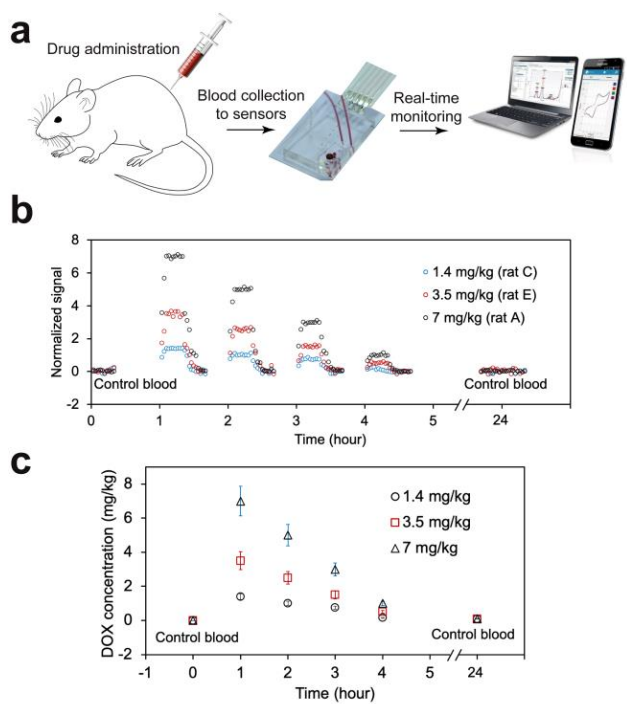

*Figure 2: Real time remote dynamic drug concentration measurement in vivo.*

## CONCLUSION

In this study, a wireless nanofluidic device integrating microfluidic sedimentation reservoirs, DLD arrays, an ICP nanofluidic component, and e-aptamer sensors for real-time continuous bioassays without uploading additional clean buffer solutions was successfully developed. The plasma from unprocessed whole blood is extracted by microfluidic sedimentation reservoirs and DLD arrays. Aptamers are customizable oligonucleotide sequences that can bind to target drug molecules. Upon binding with drug molecules, the aptamer molecule configuration changes leading to the electrical signal change for detections. This process is reversible by applying a washing solution (i.e., generating by ICP) to remove the drug molecules from the aptamer complex. In this way, the real-time monitoring of drug concentrations *in vivo* is approached.

**ACKNOWLEDGEMENTS** We gratefully acknowledge the funding provided by the City University of Hong Kong (9610467), the National Research Foundation Singapore (NRF, SBP), the National Medical Research Council Singapore, (NMRC, OFIRG) and the Ministry of Edu- cation (MOE) Singapore, Tier-2.

**CONTACT** * CHC: chiachen@cityu.edu.hk; DTP: phandinhtuan@outlook.com